# Failure Mode Analysis in Microsecond UV Laser Annealing of Cu Thin Films

Remi Demoulin
*LAAS-CNRS*
Toulouse, France
rdemoulin@laas.fr

Richard Daubriac
*LAAS-CNRS*
Toulouse, France
rdaubriac@laas.fr

Louis Thuries
*Laser Systems & Solutions of Europe (LASSE)*
Gennevilliers, France
louis.thuries@screen-lasse.com

Emmanuel Scheid
*LAAS-CNRS*
Toulouse, France
scheid@laas.fr

Fabien Rozé
*Laser Systems & Solutions of Europe (LASSE)*
Gennevilliers, France
fabien.roze@screen-lasse.com

Fuccio Cristiano
*LAAS-CNRS*
Toulouse, France
cfuccio@laas.fr

Toshiyuki Tabata
*Laser Systems & Solutions of Europe (LASSE)*
Gennvilliers, France
toshiyuki.tabata@screen-lasse.com

Fulvio Mazzamuto
*Laser Systems & Solutions of Europe (LASSE)*
Gennevilliers, France
fulvio.mazzamuto@screen-lasse.com

*Abstract*—The need of surface-localized thermal processing is strongly increasing especially w.r.t three-dimensionally (3D) integrated electrical devices. UV laser annealing (UV-LA) technology well addresses this challenge. Particularly UV-LA can reduce resistivity by enlarging metallic grains in lines or thin films, irradiating only the interconnects for short timescales. However, the risk of failure in electrical performance must be correctly managed, and that of UV-LA has not been deeply studied yet. In this work microsecond-scale UV-LA is applied on a stack comparable to an interconnect structure (dielectric/Cu/Ta/SiO$_2$/Si) in either melt or sub-melt regime for grain growth. The failure modes such as (i) Cu diffusion into SiO$_2$, (ii) O incorporation into Cu, and (iii) intermixing between Cu and Ta are investigated.

## I. INTRODUCTION

Recently, UV-LA has started to be applied in advanced interconnects so that the line resistivity can be reduced thanks to short-timescale high-temperature processing [1-2]. Efforts to study such processes is continuously rising because of rapidly growing interest from the industry on 3D-integrated electrical devices, where the applicable thermal budget may be further restricted.

With UV-LA, two different regimes are possible: (I) melt and (II) sub-melt. The melt regime is interesting because it may allow huge grain growth from liquid metals. Indeed, we already demonstrated micrometer-scale grain growth in blanket Cu thin films [3]. If this regime can be properly applied in patterned lines without introducing any negative side effects such as structural destruction, metal oxidation, and surface roughness increase, quasi grain-boundary-free lines become possible, mitigating the exponential resistivity increase in narrow lines. On the other hand, the sub-melt regime, while producing a less pronounced resistivity gain w.r.t. to the melt regime, can considerably minimize the drawbacks that are detrimental for industrial perspective. In addition, the sub-melt process is the only option for metals alternative to Cu such as Ru [4], Mo [4], and binary alloys (NiAl [5]). For them, the melting point is too high to be reached in BEOL, even for a surface selective process such as UV-LA. Still, in the sub-melt regime, large Cu grain growth is achievable. Indeed, the mean grain size observed in 50-nm-thick Cu film could go up to 414 nm after microsecond-scale sub-melt UV-LA [3], whereas it is typically about 100 nm after 600 °C furnace anneal [6-7].

Nevertheless, the risk of failure in electrical performance must be correctly assessed also in the sub-melt regime. Possible failure modes are (i) Cu diffusion into SiO$_2$, (ii) O incorporation into Cu, and (iii) intermixing between Cu and Ta. To investigate them, in this work, microsecond-scale UV-LA is applied to a stack of dielectric/Cu/Ta/SiO$_2$/Si in either melt or sub-melt regime for grain growth.

## II. EXPERIMENTAL

A 47-nm-thick sputtered-Cu was deposited on a 11 nm Ta/100 nm SiO$_2$/Si stack. A 7-nm-thick Si-based dielectric capping layer was deposited by plasma deposition at 100 °C for 10 s. Microsecond-scale UV-LA was performed at room temperature (RT) under a N$_2$ flow. Fig. 1 shows the simulated time-temperature profiles in Cu for the applied UV-LA conditions. Processes A and B kept the Cu layer in solid state, whereas Process C fully melted it. These conditions had a significant impact, particularly resulting in the reduction of sheet resistance (Fig. 2) compared to the non-annealed stack (0.60 ohm/sq): Process A gave 0.51 ohm/sq (-15%), B 0.50 ohm/sq (-16%) and C 0.54 ohm/sq (-10%). The area-weighted mean grain size became larger after Process B (210 nm) than prior to UV-LA (70 nm) as reported before [3]. Five-times (x5) processing was also tested for Processes B and C to capture more clearly the possible degradation phenomena (Fig. 2). Each sample was analyzed by scanning TEM (STEM) using different techniques: High-Angle Annular Dark-Field (HAADF), Energy-Dispersive X-ray spectroscopy (EDX), and Electron Back-Scattering Diffraction (EBSD).

## III. Results and discussion

Firstly, Cu diffusion into $SiO_2$ was investigated. Fig. 3 shows the STEM-HAADF and EDX Cu-K mapping images taken after UV-LA. Process A shows no modification of the Cu layer. Process B clearly presents Cu diffusion into $SiO_2$, which is reinforced by increasing the number of process steps to x5. Interestingly, Process "C x5", corresponding to the melt regime (cf. Fig. 1) exhibits no Cu diffusion, although the applied thermal budget is higher than Process "B x5". In fact, it has been shown that, during thermal process in the solid phase, Cu diffusion paths can be opened within the Ta layer at temperatures around 800 °C in the case of furnace anneals (a higher temperature might be required when shortening the process timescale) [8]. We speculate that the difference of the coefficient of thermal expansion between Cu (~20 × $10^{-6}$ /°C at 0-1000 °C [9]) and Ta (~7 × $10^{-6}$ /°C at 27-2400 °C [10]) introduces a tensile strain in the Ta layer during the anneal, leading to its fracture when the mechanical fracture limit is exceeded. The high-resolution TEM observations reported in Ref. 8 also support such scenario. In contrast, when fully melting the Cu layer, tensile deformation of the Ta layer can no longer occur. No Cu is therefore expected to diffuse into the $SiO_2$ layer, in agreement with the STEM-EDX observations. The diffusion barrier ability can be improved by partially (or fully) replacing Ta with TaN [11].

Secondly, O incorporation into the Cu layer was investigated. Fig. 3 shows the STEM-HAADF and EDX O-K maps taken after UV-LA. The intensity of the O signal within the Cu layer appears to increase when increasing the sub-melt thermal budget, with a concentration gradient going from the bottom of the Cu layer up to the surface. In addition, an O segregation peak is observed at the Cu/Ta interface, which also increases with the sub-melt thermal budget. These observations clearly suggest that the underlying $SiO_2$ layer is the source of the observed O, whose diffusion into the Cu layer is allowed by the mechanical rupture of the Ta layer. Indeed, in the case of Process "C x5", Ta layer fracture does not occur, which avoids any O incorporation into the Cu layer, in agreement with the STEM-EDX observations (cf. Fig. 3). The O profile is comparable between Processes A and "C x5". Based on the observed shape of the O EDX profile and other process related parameters (such as the process temperature or the thickness of the dielectric capping layer), all other possible sources of O can be excluded, including the annealing ambient, the capping layer itself or a possible residual O content in the as-deposited Cu layer.

Thirdly, intermixing between the Cu and Ta layers was investigated. Fig. 4(a) shows the EDX Ta-M signal line scans taken after UV-LA across the whole structure. In fact, the Ta-M X-ray energy is very close to the Si-K one. The two signals are therefore overlapped, however, due to the expected absence of Si in the Cu and Ta layers, the recorded signal in the top 60 nm layer can be safely related to Ta, while the deeper part corresponds to the Si contained in the $SiO_2$ layer and the Si substrate (the latter being used as a normalization reference). A zoomed view of the Ta layer region is presented in Fig. 4(b). Process A shows the as-deposited Ta thickness. For Processes B and "B x5", a weak (but not negligible) Ta penetration into the copper layer is clearly observed, due to Cu/Ta interdiffusion in the solid phase, which increases with the thermal budget. For Process "B x5", an "intermediate" Cu/Ta intermixing layer is also visible by TEM (Fig. 4(c)). In contrast, in the case of Process "C x5", the diffusion of Ta into the Cu layer is accompanied by a strong degradation of the Ta/Cu interface (Fig. 4(d)), which is probably due to the enhanced diffusion of liquid Cu (the melting point is at 1084.6 °C [12]) into the solid Ta layer (the melting point is at 3016.8 °C [12]). We speculate that the increase of electron scattering associated to the degraded Cu/Ta interface is partially responsible for the observed increase of sheet resistance in this case (Fig. 2). Finally, Ta segregation in liquid Cu is expected to result in Ta accumulation at the surface at the end of the solidification process, in agreement with the strong surface Ta peak observed by EDX for Process "C x5" (Fig. 4(a)). This could also be an additional source of electron scattering and contribute to the increase of sheet resistance (Fig. 2).

## IV. Conclusion

The dielectric/Cu/Ta/$SiO_2$/Si structure processed by microsecond UV-LA was deeply investigated to assess the risk of failure in electronic performance. The results clearly showed that it is possible to achieve a resistivity improvement in the sub-melt regime (Process A) without any drawback such as the Cu diffusion into $SiO_2$, the O diffusion from $SiO_2$ to Cu, and the inter-diffusion or interface roughening occurring at the Cu/Ta interface. Replacing Ta with a more thermally stable material in contact with Cu will help to extend a UV-LA process window in Cu-based BEOL interconnects.

## Acknowledgment

This project has received funding from the ECSEL Joint Undertaking (JU) under grant agreement No 875999. The JU receives support from the European Union's Horizon 2020 research and innovation programme and Netherlands, Belgium, Germany, France, Austria, Hungary, United Kingdom, Romania, Israel. The project has also received funding from the French PSPC project IT2-FR.

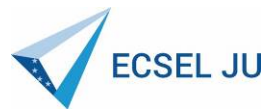
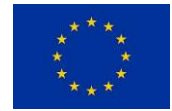

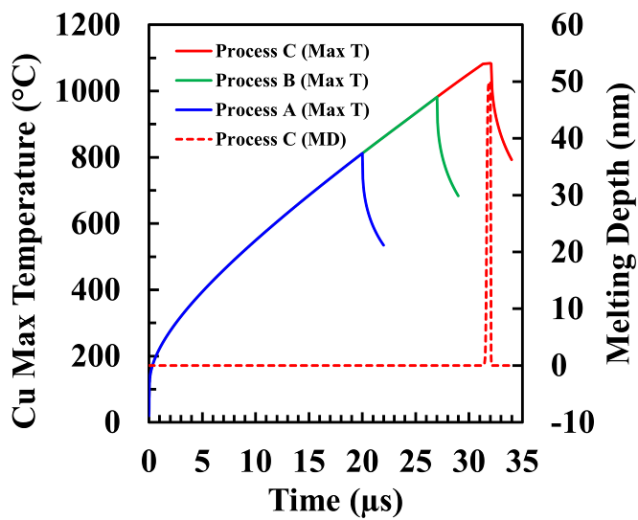

**Fig. 1.** Time vs. Cu maximum temperature (Max T) and melting depth (MD) profiles obtained by process simulation for the applied UV-LA conditions.

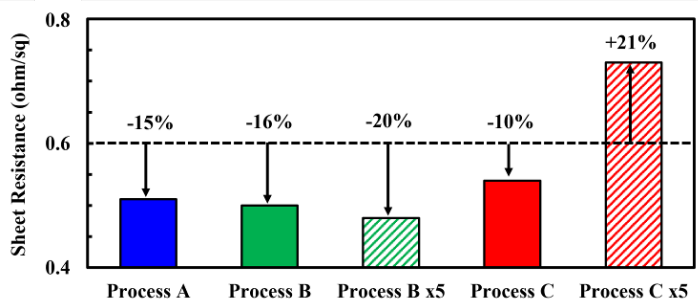

**Fig. 2.** Sheet resistance of the non-annealed (shown by the dotted line at 0.60 ohm/sq) and UV-LA samples.

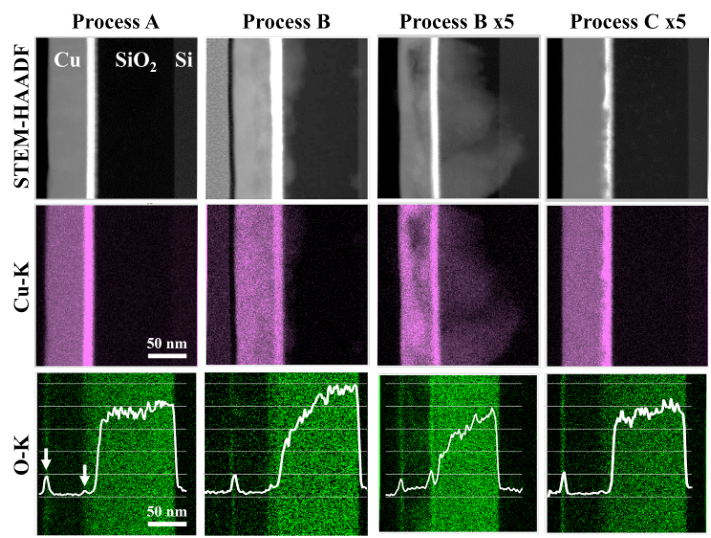

**Fig. 3.** STEM-HAADF, EDX Cu-K mapping, and EDX O-K mapping images taken for the applied UV-LA conditions. The O composition profiles are also shown on the EDX images.

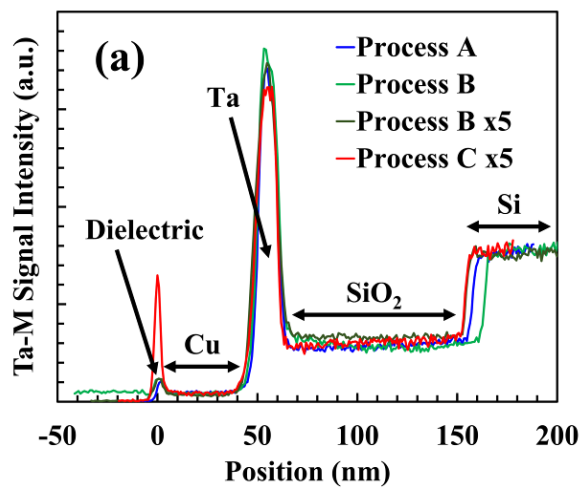

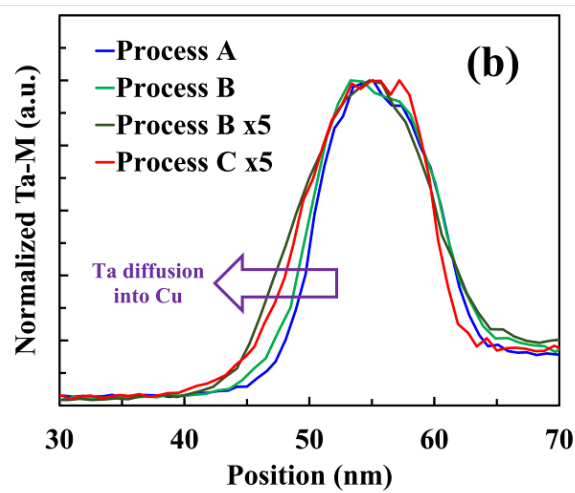

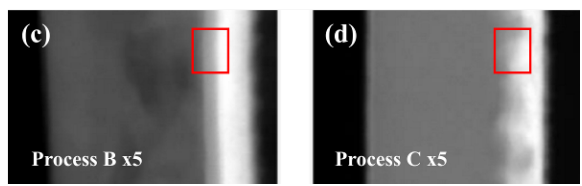

**Fig. 4.** (a) EDX Ta-M line scan profiles taken across the whole structure for the applied UV-LA conditions. It should be noted that the Si-K and Ta-M signals are overlapped. The signal intensity is roughly calibrated in Si and $SiO_2$. (b) The same profiles but normalized by the peak representing Ta between Cu and $SiO_2$, showing significant Ta diffusion into the bottom of the Cu layer after Process B x5 and C x5. (c) (d) STEM-HAADF images taken for these UV-LA conditions, showing some contrast change within the Cu/Ta interface.